\documentclass[a4paper]{llncs}

\usepackage{amsfonts}
\usepackage{hyperref}
\hypersetup{
  pdftitle={mizar-items: Exploring fine-grained dependencies in the Mizar Mathematical Library},
  pdfauthor={Jesse Alama},
  pdfkeywords={mizar, dependency graphs, proof analysis, reverse engineering, reverse mathematics},
  colorlinks=true,
  linkcolor=black,
  citecolor=black,
}
\usepackage{url}
\usepackage[utf8]{inputenc}
\usepackage{xspace}
\usepackage{xcolor}

\def\systemname#1{\textsf{#1}\xspace}
\def\libname#1{\textsf{#1}\xspace}
\newcommand{\mizar}{\systemname{Mizar}}
\newcommand{\mizaritems}[0]{\systemname{mizar-items}}

\newcommand{\MML}{\libname{MML}}

\title{\mizaritems: Exploring fine-grained dependencies in the \mizar{} Mathematical Library}
\titlerunning{\mizaritems}
\author{Jesse Alama\inst{1}\thanks{The author was supported by the FCT project
    ``Dialogical Foundations of Semantics'' (DiFoS) in the ESF
    EuroCoRes programme LogICCC (FCT LogICCC/0001/2007).  The author thanks Artur Korn\l{}owicz, Karol P\c{a}k and Josef Urban for offering their \mizar{} expertise.  The final publication is available at \href{http://dx.doi.org/10.1007/978-3-642-22673-1_19}{www.springerlink.com}.}}
\authorrunning{Alama}
\institute{Center for Artificial Intelligence\\New University of Lisbon\\\email{j.alama@fct.unl.pt}}
\begin{document}

\maketitle

The \MML{} is one of the largest collection of formalized mathematical
knowledge that has been developed with various interactive proof
assistants.  It comprises more than 1100 ``articles'' summing to
nearly 2.5 million lines of text, each consisting of a unified
collection of mathematical definitions and proofs. Semantically, it
contains more than 50000 theorems and more than 10000 definitions
expressed using more than 7000 symbols.  It thus offers a fascinating
corpus on which one could carry out a number of experiments.  This
note discusses a system for computing fine-grained dependencies among
the contents of the \MML.  For an overview of \mizar{},
see~\cite{mizar-in-a-nutshell}; for a discussion of some successful
initial experiments carried out with the help of \mizaritems,
see~\cite{premise-selection,alama-mamane-urban}.

We say that a definition, or a theorem, $\phi$ \emph{depends} on some
definition, lemma or other theorem $\psi$, (or equivalently, that
$\psi$ is a \emph{dependency} of $\phi$) if $\phi$ ``needs'' $\psi$ to
exist or hold.  The main way such a ``need'' arises is that the
well-formedness or the justification of provability does not hold in
the absence of $\psi$.  Other senses of mathematical ``dependency''
are available that are related to what we describe here, but which
\mizaritems{} does not treat (at present).  One might be interested,
for example, in the space of all \emph{logically possible} proofs of a
certain result.  Our interest is, to put it philosophically,
intensional rather than extensional: we are interested in computing
what minimally accounts for the success of a \emph{specific}
mathematical proof that has been formalized in the \mizar language.
The extensional problem is what we are after, in the long run, but
since we must work with specific formalizations of mathematical
knowledge, we need to take an intensional approach.

The primary motivation behind \mizaritems{} was the lack of a tool for
giving a complete answer to the question of what, precisely, a
\mizar{} text depends upon.  This turns out to be rather non-trivial
task.  The difficulty stems primarily from various mechanisms (such as
type inference) for making \mizar{} texts ``smoother'' for the author
and human consumer because these mechanisms, by suppressing
inferences---sometimes trivial, other times mathematically
significant---can be ``exposed'' only through much computation.

Naturally, not all items in the vast \mizar{} library are equally
interesting.  \mizaritems{} and its accompanying website (see below)
was motivated by the problem of discovering dependency information not
for arbitrary \mizar{} items, but specifically for those with
substantial mathematical or foundational value, such as the Jordan
curve theorem, the axiom of choice, the existence of strongly
inaccessible cardinal numbers, or Euler's polyhedron formula (to name
only a biased handful of examples).  The fine-grained
dependency data exposed by \mizaritems{} could also be used in theory
exploration and reverse mathematics~\cite{simpson-sosoa} or
Lakatos-style~\cite{proofs-and-refutations} investigations of
necessary and sufficient conditions for mathematical theorems.

We compute the fine-grained dependency graph for the \MML{} by
starting with an over-approximation of what is known to be sufficient
for an item to be \mizar{}-verifiable and then successively refining
this over-approximation toward a minimal set of sufficient conditions.
The method can be fairly characterized as brute-force: for each
\mizar{} item, we successively hide implicit information normally
kept hidden from a human \mizar{} formalizer, then see whether \mizar{} can still verify it.  It turns out
that this approach is rather slow; we needed to develop various
heuristics to make the brute-force computation smarter.

\mizaritems{} is accompanied by a website,
\begin{center}
  \url{http://mizar.cs.ualberta.ca/mizar-items/}
\end{center}
for exploring these dependencies.  With the site one can view any
particular \mizar{} item and see precisely what it depends upon (and
what depends on the item).  With the dependency graph, one can
ask such queries as: \emph{Is there a path between two given items?} \emph{Do all
paths from one item to another pass through a given intermediate node?}
\emph{Are there any paths between two given items that do not pass through a
given node?}

To facilitate exploration, one can start by visiting a list of
selected interesting entry points into the vast \mizar{} library.

\mizaritems{} is a collection of programs in Common Lisp, Perl,
Pascal, as well as shell scripts.  The code is available online at
\begin{center}
  \url{https://github.com/jessealama/mizar-items}
\end{center}
(The Pascal sources are not included here: They are part of the
\mizar{} code base, which, at present, is not publicly available.)

\bibliographystyle{splncs03}
\bibliography{alama}
\end{document}